\documentclass{PoS}
\usepackage{amsmath}
\usepackage[utf8]{inputenc}

\title{Traps in hadron spectroscopy:\\ Thresholds, triangle singularities, 
\ldots}

\ShortTitle{Traps in hadron spectroscopy}

\author{\speaker{Feng-Kun Guo}
\thanks{I would like to thank all my collaborators for sharing their insights into the topics discussed here. This work is supported in part by NSFC and DFG through funds provided to the Sino--German Collaborative Research Center ``Symmetries and the Emergence of Structure in QCD'' (NSFC Grant No.~11621131001, DFG Grant No.~TRR110), by NSFC (Grant No.~11647601), by the
Thousand Talents Plan for Young Professionals, and by the CAS Key Research Program of Frontier Sciences (Grant No.~QYZDB-SSW-SYS013).}\\
        CAS Key Laboratory of Theoretical Physics,
            Institute of Theoretical Physics,\\ Chinese Academy of Sciences,
            Beijing 100190, China\\
        School of Physical Sciences,
            University of Chinese Academy of Sciences,
            Beijing 100049, China\\
        E-mail: \email{fkguo@itp.ac.cn}}

\abstract{The knowledge of hadron spectrum is based on experimental observations of hadronic resonances. The resonances are normally observed as peaks in certain invariant mass distributions. However, neither is a peak necessarily due to the presence of a resonance, nor does a resonance necessarily lead to a peak. Kinematic singularities can also produce peaks. Here, we discuss such possibilities and methods distinguishing genuine resonances from kinematic effects. }

\FullConference{XVII International Conference on Hadron Spectroscopy and Structure - Hadron2017\\
		25-29 September, 2017\\
		University of Salamanca, Salamanca, Spain}

\begin{document}

\section{Introduction}

Color confinement forces us to understand hadron spectroscopy in order to understand the strong interaction at low energies. The empirical knowledge of hadron spectroscopy is provided by experimental observations of hadronic resonances and  measuring their properties. Most of the resonances were observed as peaking structures in certain invariant mass distributions, such as the new hadronic(-like) structures observed since 2003 in high energy experiments BaBar, Belle, BESIII, LHCb etc. Many of these new structures do not fit in the expectations of quark models treating mesons and baryons as quark-antiquark and three-quark bound states, respectively. Thus, they are regarded as prominent candidates of exotic hadrons which are expected to exist in the spectrum of quantum chromodynamics (QCD) as well and have not received unambiguous experimental confirmation. Most of these discoveries were made in the heavy-flavor, in particular the heavy quarkonium (the so-called $XYZ$ states), sector. For recent reviews, we refer to Refs.~\cite{Chen:2016qju,Chen:2016spr,Lebed:2016hpi,
Esposito:2016noz,Guo:2017jvc,Ali:2017jda,Olsen:2017bmm}.

However, it is well-known that resonances do not always appear as peaks. Depending on the presence of coupled channels and/or the interference with background contributions, a resonance may even show up as a dip, see, {\it e.g.}, Ref.~\cite{Taylor}. Similarly, not all peaks are due to resonances. Here, by resonances we refer to poles of the $S$-matrix. They are  of dynamical origin in the sense that they are generated as poles in the scattering amplitudes by the interactions among quarks and gluons (or among hadrons). This is necessarily a nonperturbative
phenomenon. In addition to the dynamical poles, the $S$-matrix also has kinematic singularities. The simplest is the two-body branch points (and the associated cuts) at normal thresholds. A more complicated type is the so-called triangle singularity originating from three on-shell intermediate particles.
They emerge in the physical amplitudes and can have observable effects when
the kinematics of a process satisfies special conditions.\footnote{For detailed discussions about the triangle singularity and other Landau singularities, we refer to the monographs~\cite{Eden:book,Chang:book,Gribov:book,Anisovich:2013gha} and recent lecture notes~\cite{Aitchison:2015jxa}.} Sometimes, such kinematic singularities may produce peaks mimicking the behavior of a resonance. They lay traps in hadron spectroscopy. In order to establish an unambiguous hadron spectroscopy,  it is thus important to distinguish  kinematic singularities from genuine resonances.

\section{Two-body threshold cusps}

Denoting the amplitude for producing a pair of particles with masses $m_1$ and $m_2$ in a process as $F(s)$, the two-body unitarity requires 
\begin{equation}
  \text{Im}\, F(s) = T^*(s) \rho(s) F(s) \theta(s-(m_1+m_2)^2),
\end{equation}
where $T(s)$ is the scattering amplitude, and $\rho(s) = \sqrt{\lambda(s,m_1^2,m_2^2)}/(16\pi s)$, with $\lambda(x,y,z)=x^2+y^2+z^2-2xy-2yz-2zx$ the K\"all\'en function, is the two-body phase space factor. One sees that at the threshold, there is a square-root branch point, which leads to a cusp at an $S$-wave threshold.\footnote{For higher partial waves, the cusp is smeared by positive powers of momentum in $T(s)$.} Since the location and involved hadron masses are fixed, the shape of the cusp measures the interaction strength. A well-known example is provided by the precise measurement of the $\pi\pi$ $S$-wave scattering length from the cusp at the $\pi^+\pi^-$ threshold (discussed first in Ref.~\cite{Meissner:1997fa}) in the $\pi^0\pi^0$ invariant mass distribution of the $K^\pm\to \pi^\pm\pi^0\pi^0$ data by the NA48/2 Collaboration~\cite{Batley:2000zz} (see, {\it e.g.},
Refs.~\cite{Cabibbo:2004gq,Gasser:2011ju} for theoretical discussions). The cusp in this process is moderate because the $\pi\pi$ low-energy interaction is rather weak due to the chiral symmetry breaking of QCD.

Some of the new $XYZ$ states are located close to certain $S$-wave thresholds. For instance, the $X(3872)$~\cite{Choi:2003ue} and $Z_c(3900)$~\cite{Ablikim:2013mio,Liu:2013dau} are very close to the $D\bar D^*$ threshold, the $Z_c(4020)$~\cite{Ablikim:2013xfr} is close to the $D^*\bar D^*$ threshold, the charged bottomonium-like $Z_b(10610)$ and $Z_b(10650)$~\cite{Belle:2011aa} are nearby the $B\bar B^*$ and $B^*\bar B^*$ thresholds, respectively, and their quantum numbers are the same as the corresponding $S$-wave meson pairs. Furthermore, all these structures have a narrow width. These facts stimulated models speculating them as threshold cusps~\cite{Bugg:2004rk,Bugg:2011jr,Chen:2011pv,Chen:2013coa,Swanson:2014tra}.  A common feature of these calculations is that they considered the processes with the $Z_{c(b)}$ structures in the inelastic channels, {\it i.e.}, in the modes with one pion and one charmonium (bottomonium) (here the channel with the relevant threshold is denoted as ``elastic''), so that the final states were produced through the $D^{(*)}\bar D^* (B^{(*)}\bar B^*)$ rescattering at the one-loop level. It seems that experimental data could be reproduced rather well by adjusting the cutoff parameter in the form factor which was introduced to regularize the ultraviolet divergent loop integral. But does this imply that the data suggest these structures to be simply due to coupled-channel threshold cusps, which would mean that there is no nearby pole in the $S$-matrix? To answer this question, one has to analyze elastic processes, as will be discussed below.

\begin{figure}[tbh] 
\begin{center}
\includegraphics[height=5.cm]{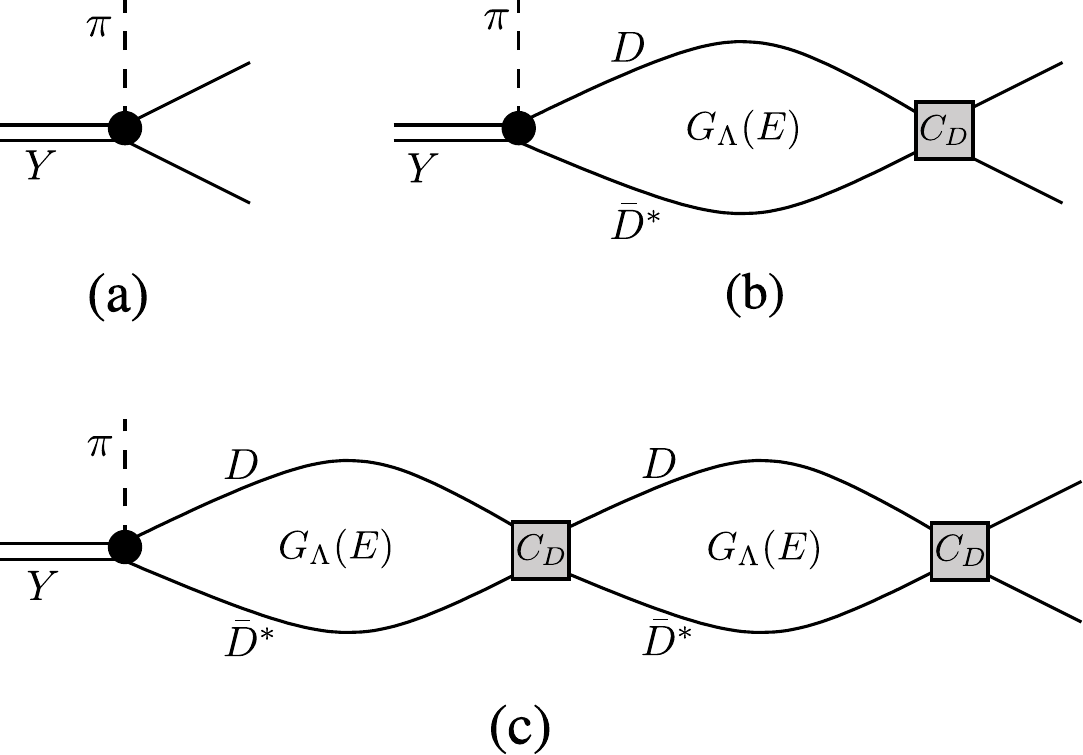}\hfill
\includegraphics[height=5.cm]{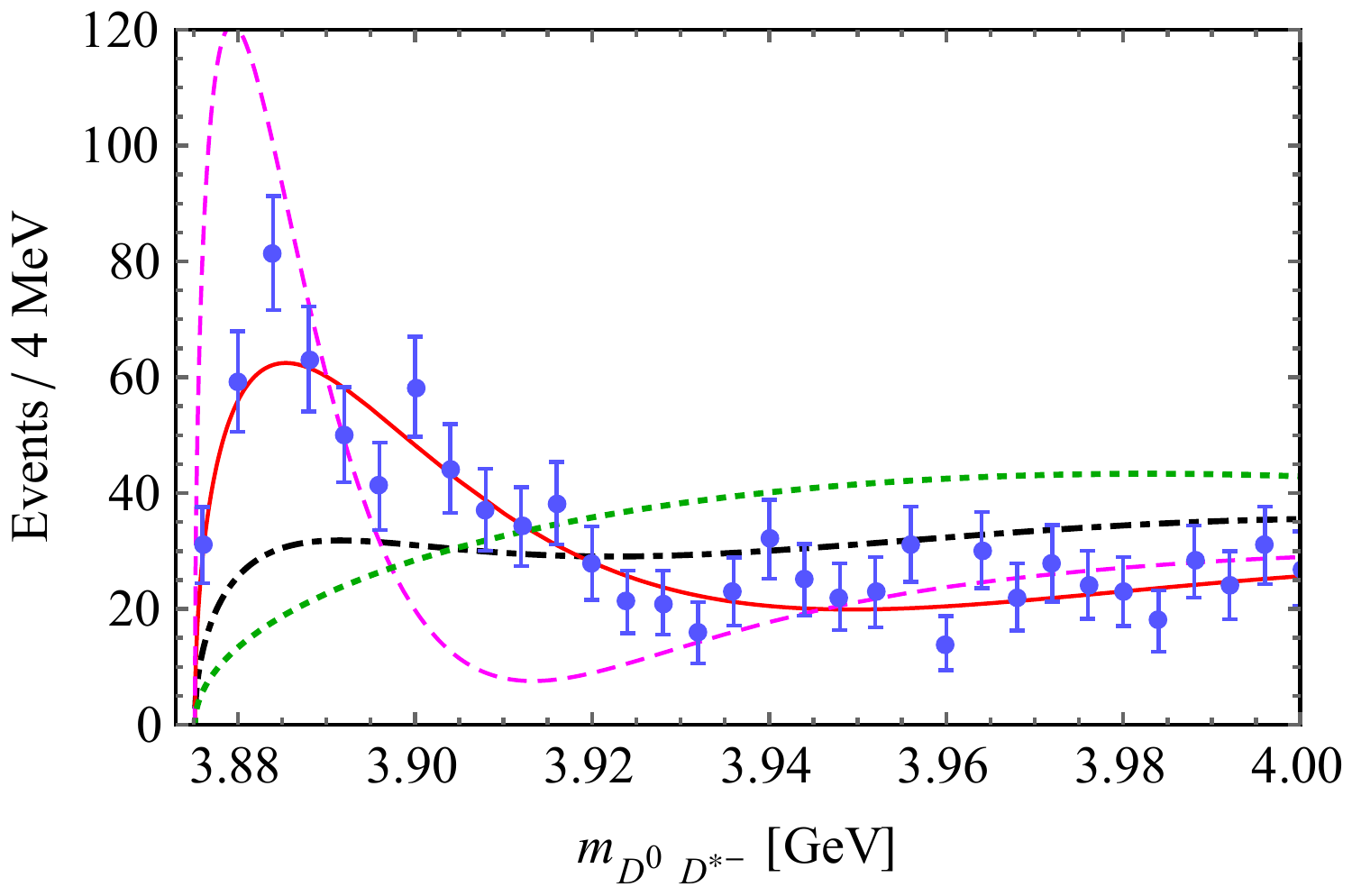}
\caption{Left: Tree-level, one-loop and two-loop diagrams for the decay $Y(4260)\to \pi D\bar D^*$. Right: Results for the $D\bar D^*$ invariant mass distribution of the decay
$Y(4260)\to \pi D \bar D^*$. The data are from Ref.~\cite{Ablikim:2013xfr} and the results from the tree level, one-loop and two-loop
calculations are shown by the dotted (green), solid (red) and dashed (magenta) curves, respectively. The parameters are determined from fitting to data using the one-loop amplitude.
The dot-dashed (black) curve shows the one-loop result with the rescattering strength requested to be small to justify a perturbative treatment.}
 \label{fig:DDst}
\end{center}
\end{figure}

The intrinsic assumption of the approaches outlined in the cusp models~\cite{Bugg:2011jr,Chen:2013coa,Swanson:2014tra} is that the interactions are perturbative so that the amplitude can be approximated by the one loop result, which does not possess a pole by definition. Let us consider a two-channel problem, say $J/\psi\pi$ and $D\bar D^*+c.c.$\footnote{The required charge conjugation will be kept implicit in the following.} We denote the production vertex for these two modes from some process as $g_\text{in}$ and $g_\text{el}$, respectively, and approximate the tree-level $S$-wave amplitudes for $J/\psi\pi\to D\bar D^*$ and $D\bar D^*\to D\bar D^*$ as constants $C_X$ and $C_D$, respectively. The direct $J/\psi\pi\to J/\psi\pi$ amplitude may be neglected due to the Okubo--Zweig--Iizuka rule. Thus, the cusp models for the production of $J/\psi\pi$ and $D\bar D^*$ may be expressed as the following one-loop amplitudes
\begin{eqnarray}
  g_\text{in} + g_\text{el}\, G_{\Lambda}(E)\, C_X, \qquad\text{and}\qquad 
  g_\text{el} \left[1 + G_{\Lambda}(E)\, C_D \right],
  \label{eq:1loop}
\end{eqnarray}
respectively, where $G_\Lambda(E)$ is the two-point loop function with $D\bar D^*$ as the intermediate states and $\Lambda$ denotes that the loop integral needs to be regularized. The two terms in the second amplitude are represented as (a) and (b) in the left panel of Fig.~\ref{fig:DDst}. 
One notices that the $G_{\Lambda}(E)\, C_D$ can be fixed from fitting to the $D\bar D^*$ invariant mass distribution in the near-threshold region (so that the approximation of the contact term as a constant $C_D$ is valid) because $g_\text{el}$ only serves as an overall normalization and does not affect the shape. On the contrary, $G_{\Lambda}(E)\, C_X$ cannot be fixed due to its interference with $g_\text{in}$. Once $G_{\Lambda}(E)\, C_D$ is fixed from fitting to the data using the one-loop amplitude, it is easy to check whether the implicit perturbative assumption is proper by comparing the two-loop, 
\begin{equation}
  g_\text{el} \left[1 + G_{\Lambda}(E)\, C_D + G_{\Lambda}(E)\, C_D\, G_{\Lambda}(E)\, C_D \right],
\label{eq:2loop}
\end{equation}
with the one-loop result. If the difference is small, the perturbative treatment is valid; otherwise, it would mean that such a model is not self-consistent.

We regularize the loop integral $G_{\Lambda}(E)$ using a Gaussian form factor. Using the one-loop amplitudes given in Eq.~\eqref{eq:1loop}, we find that indeed the BESIII data for both the $(D\bar D^*)^-$ invariant mass distribution of the $e^+e^- \to \pi^+ (D\bar D^*)^-$~\cite{Ablikim:2013xfr} and the $J/\psi\pi^-$ invariant mass distribution of the $e^+e^- \to \pi^+ \pi^-J/\psi$~\cite{Ablikim:2013mio} can be well described, both measured at the $e^+e^-$ center-of-mass energy $E_\text{cm}=4.26$~GeV. The best fit to the data for the former process is shown as the solid curve in the right panel of Fig.~\ref{fig:DDst}, in comparison with the data. 
Using the same parameters, the tree-level, which is simply the phase space, and two-loop, Eq.~\eqref{eq:2loop}, results are shown as the dotted (green) and dashed (magenta) curves, respectively. It is clear that the two-loop result largely deviates from the one-loop one, which indicates that the interaction determined in this way is nonperturbative or $ G_\Lambda(E)\,C_D$ is of order 1 in the near-threshold region. In fact, if one resums the two-point bubbles up to infinite orders, the resulting amplitude $g_\text{el} /\left[1 -G_{\Lambda}(E)\, C_D \right]$ has a pole close to the $D\bar D^*$ threshold. It is the narrowness of the near-threshold peak that requires the $D\bar D^*$ interaction to be nonperturbative. If we demand the interaction to be perturbative by hand, say by requiring $| G_\Lambda(E)\,C_D| = 1/2$ at the $D\bar D^*$ threshold, we are not able to produce any narrow structure in the $D\bar D^*$ channel and the corresponding result is shown as the dot-dashed (black) curve in the right panel of Fig.~\ref{fig:DDst}.
On the contrary, the data in the inelastic channel, $e^+e^- \to \pi^+ \pi^-J/\psi$ for the $Z_c(3900)$, is not enough to determine the rescattering strength because it cannot be disentangled from the direct production represented by $g_\text{in}$ in the first amplitude in Eq.~\eqref{eq:1loop}.\footnote{This is different from the case of $K^\pm\to \pi^\pm\pi^0\pi^0$ where the two channels $\pi^0\pi^0$ and $\pi^+\pi^-$ are related to each other.}

Therefore, we conclude that a narrow pronounced near-threshold peak cannot be produced solely by a threshold cusp, and it necessarily indicates the existence of a nearby pole which might be even a virtual state pole located in the unphysical Riemann sheet with respect to the elastic channel. In fact, it was suggested in Ref.~\cite{Guo:2013sya} and the $Z_c(3900)$ and $Z_c(4020)$ correspond to virtual state poles, which may be located a few tens of MeV below the corresponding thresholds, and a multi-channel fit using a formalism with the unitarity built in~\cite{Hanhart:2015cua,Guo:2016bjq} to the Belle data suggests the $Z_b(10610)$ to be a virtual state and the  $Z_b(10650)$ to be a resonance. Here we want to briefly comment on the lattice results by the HALQCD~\cite{Ikeda:2016zwx,Ikeda:2017mee} which suggest the $Z_c(3900)$ is a threshold cusp. In the HALQCD calculation, they derived the $\pi J/\psi$, $\rho\eta_c$ and $D\bar D^*$ coupled-channel potential from lattice with the pion mass between 410 and 700~MeV. From the Lippmann--Schwinger equation, a virtual state pole far from the physical region was found.  We will not discuss their method, but only point out that the obtained $D\bar D^*$ invariant mass is too broad to account for the BESIII double $D$-tagged data with little background at $E_\text{cm}=4.26$~GeV~\cite{Ablikim:2015swa}.

It is worthwhile to notice that in the above discussion, we have assumed that the production vertex (the $Y\to \pi D\bar D^*$ vertex $g_\text{el}$ in the considered example) does not produce a nontrivial structure. The presence of nearby triangle singularities~\cite{Wang:2013cya,Wang:2013hga} makes the problem more complicated. However, as will be discussed below, the conclusion that a narrow pronounced near-threshold peak in the elastic channel requires the presence of a nearby pole remains unchanged.

\section{Triangle singularities}

The location of a threshold cusp is fixed, but the location of a triangle singularity, which is the leading Landau singularity~\cite{Landau:1959fi} of a triangle diagram, depends crucially on the kinematics, {\it i.e.}, on the masses and momenta of the involved particles. Moreover, whether it appears close to the physical region also depends on the kinematics. Coleman and Norton showed that the triangle singularity is on the physical boundary if the process could happen classically, {\it i.e.}, all of the three intermediate particles could go on shell and all of the interaction vertices satisfied the energy-momentum conservation~\cite{Coleman:1965xm}. Triangle singularity is a logarithmic branch point, which would produce an infinite reaction rate  if it really appears in the physical region. This does never happen because at least one of the three particles must be unstable as a consequence of the on-shell condition. The finite width moves the singularity into the complex energy plane, and the differential reaction rate can have a finite peak due to the proximity of the singularity. Although there have been lots of discussions since the 1960's, no unambiguous observation of a triangle singularity was achieved in the old days. In recent years, experimental data have been collected in many more processes, and there appeared several candidates which might be explained by or contain a large contribution from triangle singularities. A prominent example is provided by the $\eta(1405)\to\pi\pi\pi$~\cite{BESIII:2012aa}. The $G$-parity of the pions and the $\eta(1405)$ are negative and positive, respectively, so that this decay breaks isospin symmetry. In Refs.~\cite{Wu:2011yx,Wu:2012pg}, it is proposed that this process can happen by coupling the initial state to the $\bar K K^*$, the $K^*$ decaying into $K\pi$ and the $K\bar K$ rescattering into $\pi\pi$. The rescattering contains both the $f_0(980)$ and $a_0(980)$. Isospin symmetry breaking is derived from the mass differences between the charged and neutral intermediate strange mesons. The kinematics of the $\eta(1405/1475)\to\pi f_0(980)$ allows a triangle singularity close to the physical region, and as a consequence the isospin breaking is tremendously enhanced.\footnote{Because of isospin braking, the neutral and charged strange meson loops cancel out below the $K^+K^-$ threshold and above the $K^0\bar K^0$ threshold so that the $f_0(980)$ peak is as narrow as about 10~MeV$\sim 2(M_{K^0}-M_{K^+})$.} 
In recent years, triangle singularities were considered in the discussion on light mesons, the $a_1(1420)$~\cite{Liu:2015taa,Ketzer:2015tqa,Aceti:2016yeb}, the $f_1(1420)$~\cite{Debastiani:2016xgg,Liu:2015taa}, and the $f_2(1810)$~\cite{Xie:2016lvs}, on exotic hadron candidates, the $Z_c(3900)$~\cite{Wang:2013cya,Wang:2013hga,Szczepaniak:2015eza,Szczepaniak:2015hya,Pilloni:2016obd,Gong:2016jzb}, the $P_c(4450)$~\cite{Guo:2015umn,Liu:2015fea,Guo:2016bkl,Bayar:2016ftu} and its hidden-strangeness analogue~\cite{Xie:2017mbe}, and the $Z_b$~\cite{Wang:2013hga,Szczepaniak:2015eza,Bondar:2016pox}, and in the baryon sector, see, {\it e.g.},~\cite{Wang:2016dtb,Debastiani:2017dlz}. Suggestions of searching for new triangle singularities in $B$ or $B_c$ decays can be found in Refs.~\cite{Liu:2017vsf,Pavao:2017kcr}. In particular, the recent BESIII observation of the fast variation of the $\psi'\pi$ distribution shapes for the $e^+e^-\to \psi'\pi^+\pi^-$ measured at different collision energies~\cite{Ablikim:2017oaf} could be a hint to the importance of triangle singularities discussed in Ref.~\cite{Liu:2014spa,Cao:2018}.

%
\begin{figure}[tb]
  \centering
    \includegraphics[width=0.4\linewidth]{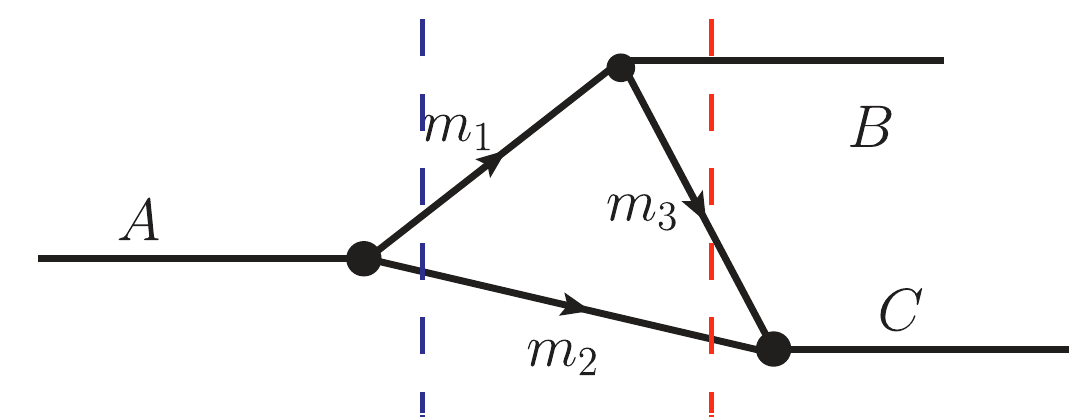}
  \caption{A triangle diagram with the internal lines labeled by the masses of the corresponding particles. The two vertical dashed lines refer to the two cuts discussed in the text.
  }
  \label{fig:triangle}
\end{figure}
To be more explicit, let us take Fig.~\ref{fig:triangle} and explain the
kinematical region where the triangle singularity occurs on the physical boundary. The diagram can be interpreted as $A$ decays into particles $m_1$ and $m_2$,
following by the sequential decay of $m_1$ into $B$ and $m_3$, and $m_2$
and $m_3$ react to generate the external $C$. Notice that $A,B$ and $C$ do not need to be single particles. We consider the rest frame of $A$. All of the intermediate particles are on shell so that the magnitudes of their momenta are fixed in terms of the $A,B,C$ invariant masses. The reactions at all vertices can happen classically means that all particles must move parallel or anti-parallel, and particle $m_3$ emitted from the decay of $m_1$ must move fast enough to catch up with particle $m_2$ in order to react to form the external $C$. Expressing the above conditions mathematically, we get~\cite{Bayar:2016ftu}
\begin{eqnarray}
  q^{}_{\rm on+} = q^{}_{a-}, \quad \text{with}\quad
   q^{}_{{\rm on}+} = \frac1{2 m^{}_A} \sqrt{\lambda(m_A^2,m_1^2,m_2^2)},~~ q^{}_{a-} = \gamma \left( \beta \, E_2^* - p_2^* \right), 
   \label{eq:ts}
\end{eqnarray}
where $E_2^*$ and $p_2^*$ are the energy and the size of the 3-momentum of particle $m_2$ in the $B$ rest frame, $\beta$ is the magnitude of the velocity of $B$ in the rest frame of $A$, and $\gamma= 1/{\sqrt{1-\beta^2}}$ is the Lorentz boost factor.  The two momenta given above correspond to the two cuts depicted in Fig.~\ref{fig:triangle}.
One sees that the triangle singularity is on the physical boundary only for very special kinematics.
For given masses $m_2$, $m_3$ and invariant masses for external particles, one can
work out the special range of $m_1$, as well as the corresponding range of the triangle singularity in, {\it e.g.}, the $C$ invariant mass. The ranges can be obtained by requiring $q_{\rm on}$ and $q_{a-}$ to take values in the physical regions.
Using the above equation, we find that when
\begin{equation}
 m_1^2 \in \left[ \frac{m_A^2 m^{}_3 + m_{B}^2 m^{}_2}{m^{}_2+m^{}_3} - m^{}_2 m^{}_3\,,~
 \left(m^{}_A-m^{}_2 \right)^2 \right],
 \label{Eq:m1range}
\end{equation}
there is a triangle singularity on the physical boundary, and in terms of the $B$ invariant mass it is within the range
\begin{equation}
 m_{C}^2 \in \left[ (m_2+m_3)^2,~ \frac{m^{}_A m_3^2 - m_{B}^2
 m^{}_2}{m^{}_A-m^{}_2} + m^{}_A m^{}_2 \right].
 \label{Eq:m23range}
\end{equation}
For discussions of these ranges, see, {\it e.g.}, Refs.~\cite{Aitchison:1964zz,Szczepaniak:2015eza,Liu:2015taa,Guo:2015umn,Guo:2016bkl}.

\begin{figure}[tb]
  \centering
    \includegraphics[height=5.cm]{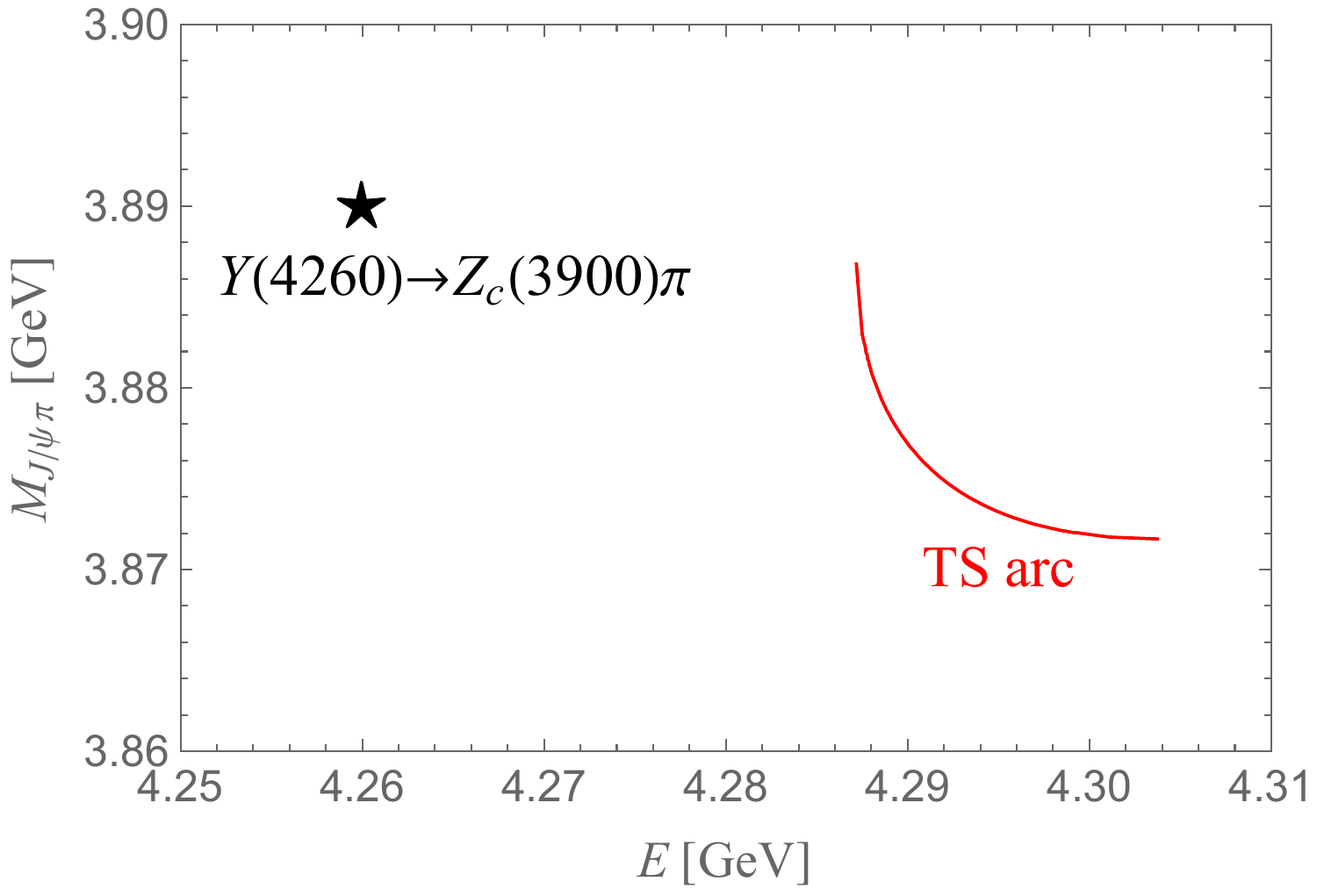} \hfill
    \includegraphics[height=4.9cm]{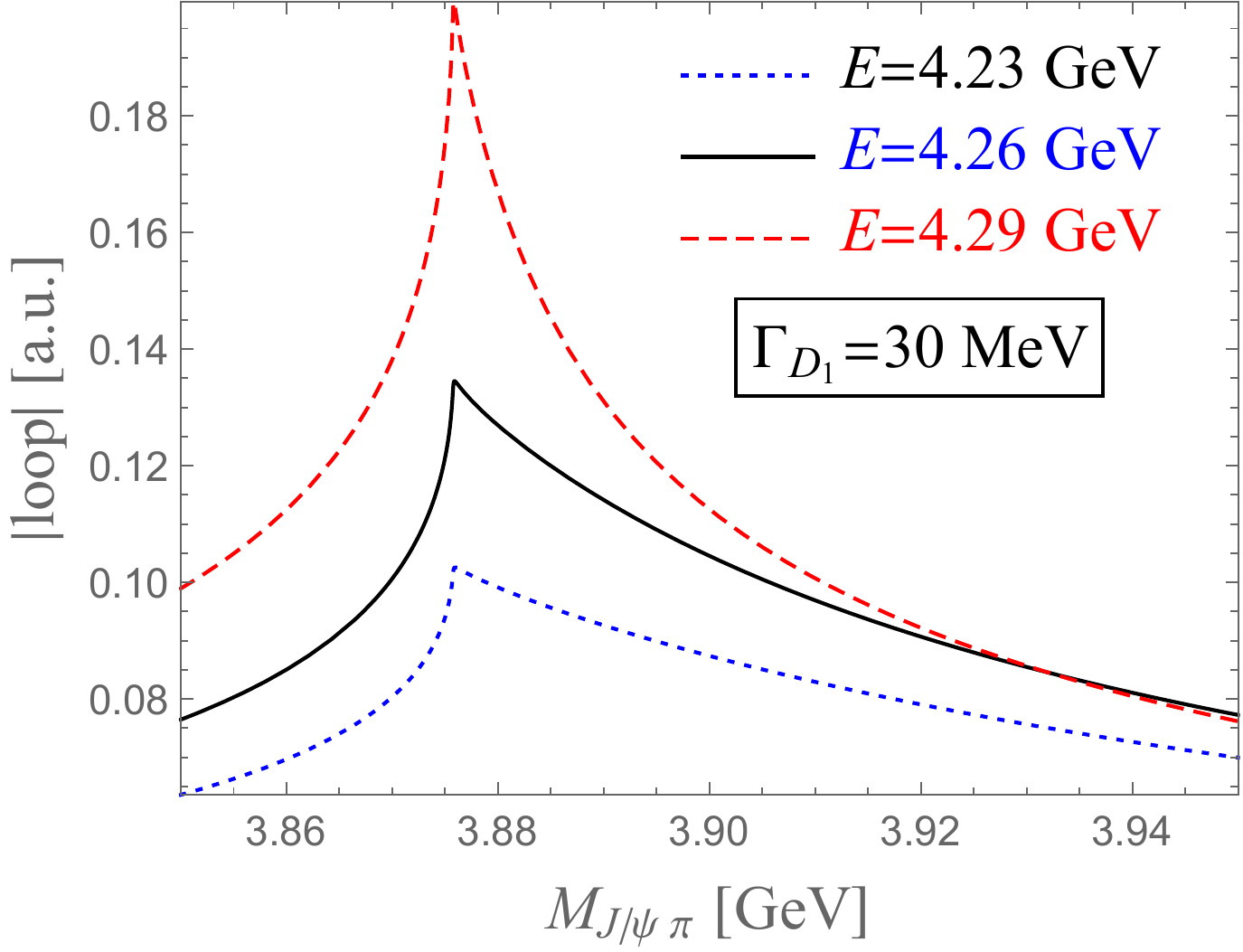}
  \caption{Left: The red cure labeled as ``TS arc'' represents the trajectory in the $E_\text{cm}$--$M_{J/\psi\pi}$ plane along which the triangle singularity is on the physical boundary. Right: Dependence of the absolute value of the $D_1\bar D D^*$ triangle loop on the incoming energy.
  }
  \label{fig:ytozpi}
\end{figure}

In order to make the sensitivity on the kinematics clear, let us take the process $e^+e^-\to J/\psi\pi\pi$, the discovery process of the $Z_c(3900)$, as an example. As first pointed out in Ref.~\cite{Wang:2013cya}, the triangle singularity of the $D_1(2420)\bar D D^*$ triangle loop (substituting $m_1,m_2,m_3, B$ and $C$ in Fig.~\ref{fig:triangle} by $D_1,\bar D, D^*, J/\psi\pi$ and $\pi$, respectively) may play an important role. If we fix the intermediate particles as $D_1\bar D D^*$ with their widths neglected and vary the collision energy $E_\text{cm}$ and the $J/\psi\pi$ invariant mass, Eq.~\eqref{eq:ts} implies that the triangle singularity appears as an arc in the $E_\text{cm}$--$M_{J/\psi\pi}$ plane as shown in the left panel of Fig.~\ref{fig:ytozpi}. The kinematics for the $Y(4260)\to Z_c(3900)\pi$ is not on the arc, but is only a few tens of MeV away and thus leaves an influence. Taking a 30~MeV constant width for the $D_1$, the absolute values of the $D_1\bar D D^*$ scalar 3-point loop integral are shown in the right panel of Fig.~\ref{fig:ytozpi}. When $E_\text{cm}=4.29$~GeV, the singularity is away from the physical region only due to the small $D_1$ width, and the loop function has a sharp peak. Decreasing $E_\text{cm}$, the peak becomes less pronounced since the triangle singularity is moving further away from the physical region. Nevertheless, there is always a cusp at the $\bar D D^*$ threshold because they couple in an $S$-wave, as discussed in the last section, and the threshold cusp is a subleading singularity of the triangle diagram. The finite width of the $D_1$ does not smear out this cusp. The sensitivity of the line shape on the incoming energy is one of the keys to reveal the role of kinematic singularities. 

The above discussion also implies that the $D_1\bar D D^*$ triangle diagrams have to be included in a realistic analysis of the $Z_c(3900)$. Such an analysis of both the $Y(4260)\to J/\psi\pi\pi$~\cite{Ablikim:2013mio} and $Y(4260)\to D\bar D^*\pi$~\cite{Ablikim:2015swa} data was done in Ref.~\cite{Albaladejo:2015lob}. It was found that despite the inclusion of the $D_1\bar D D^*$ loops, fits to the data still led to the presence of a pole corresponding to the $Z_c(3900)$ near the $D\bar D^*$ threshold. Depending on whether the $J/\psi\pi$--$D\bar D^*$ coupled-channel interaction model allows for an energy-dependent term in the potential, the pole can be either a virtual state below the $D\bar D^*$ threshold, which could be a few tens of MeV away, or a resonance above the threshold. However, a later analysis by the JPAC Collaboration using a constant $D_1 D^*\pi$ coupling concluded that the data~\cite{Ablikim:2013mio,Ablikim:2015tbp,Ablikim:2015swa,Ablikim:2015gda} did not allow for distinguishing models with a $Z_c(3900)$ state or not. It is interesting to see whether the conclusion remains if the updated BESIII data on the $e^+e^-\to J/\psi\pi^+\pi^-$ at $E_\text{cm}=4.23$ and 4.26~GeV~\cite{Collaboration:2017njt} are used and the $D_1 D^*\pi$ $D$-wave coupling is properly taken into account. In particular, one sees from Fig.~\ref{fig:ytozpi} that the triangle singularity should not be important for $E_\text{cm}=4.23$~GeV.

Another interesting occurrence of triangle singularity~\cite{Guo:2015umn,Liu:2015fea} is related to the narrow structure $P_c(4450)$, which was observed by the LHCb Collaboration~\cite{Aaij:2015tga} in the $J/\psi p$ invariant mass distribution of the decay $\Lambda_b^0\to J/\psi p K^-$. The $P_c(4450)$ is regarded as a candidate of hidden-charm pentaquark states first predicted in Ref.~\cite{Wu:2010jy}. However, it was pointed out in Ref~\cite{Guo:2015umn} that the $P_c(4450)$ mass coincides with
the $\chi_{c1}\,p$ threshold and, more importantly, the location of the triangle
singularity of the $\Lambda(1890)\chi_{c1}p$ loop diagram. The $\Lambda(1890)$ is a well-established four-star hyperon with $J^P=3/2^+$ and a width of about 100~MeV decaying with a large branching fraction into $N\bar K$~\cite{Patrignani:2016xqp}, and $J/\psi p$ in the final state are produced through the $\chi_{c1}p$ rescattering. The shape produced by the $\Lambda_b^0\to J/\psi p K^-$ scalar 3-point loop integral well reproduces the LHCb peak structure around 4.45~GeV. However, because the $\chi_{c1}p\to J/\psi p$ rescattering strength  and the $\Lambda_b\to \Lambda(1890)\chi_{c1}$ decay width are unknown and the presence of many $\Lambda^*$ resonances coupled to $p K^-$, we are not able to predict how large the triangle singularity contribution is.\footnote{For such a calculation, the three-body unitarity needs to be considered properly, which presents another difficulty.} In view of this, we need to have other methods revealing whether the $P_c(4450)$ is really a pentaquark or not, which is an utmost important question to be answered for the $P_c$'s being the first candidates of quasi-explicitly exotic pentaquark states.\footnote{The quantum numbers of the $P_c$ structures can be formed by three light quarks.
However, since their masses are above 4~GeV, if they are light baryons they would decay into light hadrons very quickly due to the vast amount of phase space, and
the widths would be much larger than those reported by the LHCb Collaboration.
Therefore, it is more natural to assume that there are a pair of charm and
anticharm quarks inside whose annihilation into light hadrons is suppressed.} Possible methods include:
\begin{itemize}
  \item To measure the $\chi_{c1}p$ invariant mass distribution of the decay $\Lambda_b^0\to \chi_{c1}p K^-$. For this process, the $\chi_{c1}p$ pair are in the final state as well as in the intermediate state. Therefore, in addition to the $\Lambda(1890)\chi_{c1}p$ loop diagram, the $\chi_{c1}p K^-$ can also be produced at tree-level by exchanging the $\Lambda(1890)$. The subtle interference between the tree-level and triangle diagrams around the singularity region results in an amplitude which is simply the tree-level one multiplied by a complex phase factor, and there would be no pronounced peak in the projected Dalitz distribution~\cite{Schmid:1967}.\footnote{Corrections to this observation were discussed in Refs.~\cite{Anisovich:1995ab,Szczepaniak:2015hya}.} Thus, were the $P_c(4450)$ due to a triangle singularity, there would be no narrow near-threshold peak in the $\chi_{c1}p$ invariant mass distribution. Following this suggestion, the LHCb Collaboration measured the branching fraction of $\Lambda_b^0\to \chi_{c1}p K^-$~\cite{Aaij:2017awb}, and the amplitude analysis is on going.
  \item To determine the quantum numbers of the $J/\psi p$ pair in the $P_c(4450)$ peak structure. For the discussed triangle singularity to produce a narrow peak, the $\chi_{c1}$ and proton need to be in an $S$-wave, and thus the quantum numbers of the rescattered $J/\psi p$ should be $J^P=1/2^+$ or $3/2^+$. So far the quantum numbers have not been unambiguously determined with the latter being one of the preferred~\cite{Jurik:2016bdm}.
  \item To search for the $P_c(4450)$ in reactions with different kinematics to avoid the discussed triangle singularity. Such reactions could be, {\it e.g.}, the photoproduction
processes~\cite{Wang:2015jsa,Kubarovsky:2015aaa,Karliner:2015voa,Huang:2016tcr},
pion induced reactions~\cite{Lu:2015fva,Liu:2016dli} and heavy ion
collisions~\cite{Wang:2016vxa,Schmidt:2016cmd}.
\end{itemize}

\section{Conclusion}

In order to understand the QCD spectrum, we need to search for more candidates of exotic hadrons. We are aware of the existence of possible traps along the way, such as resonance-like structures induced by kinematic-singularities discussed above and due to some other reasons, like the $X(5568)$ reported by the D0 Collaboration~\cite{D0:2016mwd}, which finds no reason to exist theoretically~\cite{Burns:2016gvy,Guo:2016nhb}, has no signal in lattice QCD calculations~\cite{Lang:2016jpk}, and was not confirmed in other experiments~\cite{Aaij:2016iev,Sirunyan:2017ofq,CDF:2017}. In order to escape from these traps, cooperative efforts from experiments, phenomenology and lattice calculations are necessary.

\end{document}